# The Effected Oxide Capacitor in CMOS Structure of Integrated Circuit Level 5 Micrometer Technology

S. Rodthong and B. Burapattanasiri

**Abstract**—This article is present the effected oxide capacitor in CMOS structure of integrated circuit level 5 micrometer technology. It has designed and basic structure of MOS diode. It establish with aluminum metallization layer by sputtering method, oxide insulator layer mode from silicon dioxide, $n^+$ and $p^+$ semiconductor layer, it has high capacitance concentrate. From the MOS diode structure silicon dioxide thickness 0.5 micrometer, it will get capacitance between aluminum metal layer and $p^+$ semiconductor at 28.62 pF, the capacitance between aluminum metal layer and $n^+$ semiconductor at 29.55 pF. In this article establish second metal layer for measurement density values of first aluminum metal layer with second aluminum metal layer,   it has density values at 16 pF.

—————————— ◆ ——————————

## 1 INTRODUCTION

THIS time integrated circuit technology is continuing to development, especially integrated circuit size, it has development to smaller and the contraction of p, n semiconductor is depths less 1 micrometer.  In CMOS integrated circuit technology has smaller size too, and it able to use in multiuse full especially in lowest frequency, the CMOS is not operate as much as it can, because of the big problem is the structure of MOS, it look like capacitor inside, so the hidden capacitor of MOS has important part is oxide capacitance: $C_{ox}$ (The oxide capacitance will stable depend on oxide thickness and surface area), and depletion capacitance: $C_d$ (The depletion capacitance will chance as follow bias voltage at gate). The both of circuit are including the serial. Normally, when MOS is receive bias voltage and accumulation range, the MOS capacitance equation to oxide capacitance layer and the maximum capacitance when the MOS in depletion range, summation capacitance equation to summation of oxide capacitance. Depletion capacitance including the serial, thus summation capacitance is decrease until into strong inversion range it will make the lowest summation capacitance. When passing this range, in high frequency case the summation capacitance is stable, but in low frequency case the summation capacitance is increase again, if hidden capacitance within MOS is over high it will make MOS operating lower and performance of MOS is not completely. In this article we are interested in this problem especially oxide capacitor, it is stable and not depend on voltage at gate, thus it able to decrease hidden capacitance in this part by establish processing for control hidden capacitance within oxide insulators at decrease gate.

## 2 MOS DIODE STRUCTURE

The important structure of MOS diode has three parts. First aluminum metal film, it has compound with 2% silicon (Al-Si (2wt %)) by weight, establish by sputter, metal film thickness 500 nanometer.

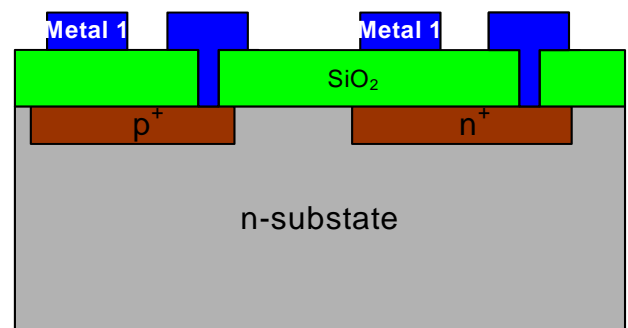

Fig. 1. MOS Diode Structure.

From figure 1 oxide capacitance, it has stable, unchanged, but depend on oxide layer thickness and contraction area is able to show the relation of capacitor and the capacitance is not change with voltage capacitor.

The capacitance is unchanged from voltage increasing or voltage decreasing, but capacitance factor is silicon dioxide layer thickness, if it high thickness then decreased capacitance, if it lower thickness then increased capacitance, after that it able to show the relation as equation (1)

$$C = \frac{\varepsilon_{ox} A}{t_{ox}} \qquad (1)$$

————————————————
- *Songpol Rodthong is with the Department of Electronic and Telecommunication Engineering, Faculty of Engineering, Kasem Bundit University, Bangkok, Thailand 10250..*
- *Bancha Burapattanasiri is with the Department of Electronic and Telecommunication Engineering, Faculty of Engineering, Kasem Bundit University, Bangkok, Thailand 10250..*

Setting A is capacitance area.
$t_{ox}$ is silicon dioxide layer thickness.
$\varepsilon_{ox}$ is permittivity of an oxide ($8.85 \times 10^{-14}$)

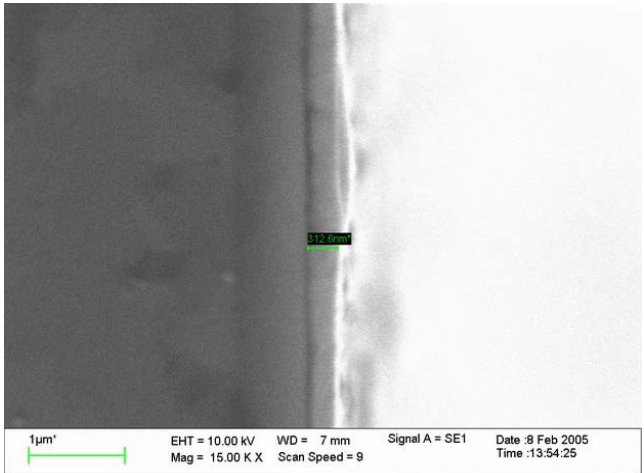

Fig. 2. Transversal image metal film layers according to the sputter condition.

From figure 2 when it receiving bias voltage it will has electro potential spread abreast in the area. Second, oxide insulator has silicon dioxide layer thickness 500 nanometers by thermal oxidation. Third, semiconductor or substrates have added impure substances for change to $n^+$ and $p^+$ semiconductor by diffusion.

## 3 DEEPLY BETWEEN P$^+$ SEMICONDUCTOR BOUNDARIES WITH N-SUBSTATE

If in MOS diode establish processing, when first mask opened completely after that high concentrate impure substance diffusion into silicon wafer. In this step we have two silicon wafer diffusion testing. Figure 3 junctions deeply p$^+$ impure substance and n-substrate

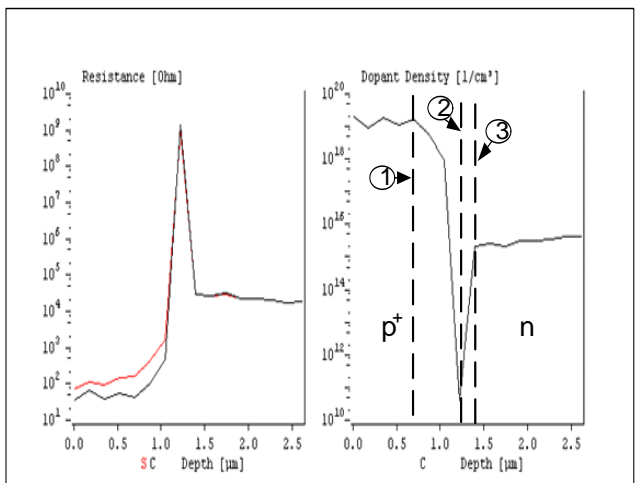

Fig. 3. Junction deeply p$^+$ impure substance and n-substrate.

First silicon wafer is for example. Second sheet is open mask already it uses for testing. We use boron diffusion n-substance, after that appear p$^+$ semiconductor on n-substance, take example silicon wafer it complete diffusion according to same condition like a real silicon wafer. When you measurement co-ordinates voltage you will get junction deeply and an impure substance density rate too as figure 3. From figure 3 junctions deeply p$^+$ impure substance and n-substrate at 1$^{st}$ point is the beginning of changing graph and able to read 0.65 micrometer, but 2$^{nd}$ is minimum point 1.25 micrometer, and 3$^{rd}$ point from graph is 1.45 micrometer. When you bring the differential between 1$^{st}$ point and 3$^{rd}$ point to minus after that p$^+$ semiconductor deep 0.8 micrometer, and density of p$^+$ impure substance dope from graph on surface $10^{19}$ atom.

## 4 CAPACITANCE AND SILICON DIOXIDE RELATIONS BETWEEN ALUMINUM METAL WITH P$^+$ SEMICONDUCTOR

From figure 4 is capacitance between aluminum metal layer and p$^+$ semiconductor by electric meter. Electric meter is able to measurement electric characteristic of circuit and electronics devices, for the functional is electric meter will send voltage pass to capacitor and show the result on monitor. The electric meter component with HP4156B precision semiconductor parameter analyzer, HP 4284A precision LCR meter and HP E5250A low leakage switch main frame. HP 4156B operating to current and voltage measurement. When send voltage increase from -5V until to 5V then measurement capacitance and voltage (C-V) as figure 4. The result from graph show measurement capacitance between aluminum metal and p$^+$ semiconductor is capacitance alteration follow on voltage. From graph that show when sanded voltage to C increase with capacitance and the measurement capacitance has be similar to all of them.

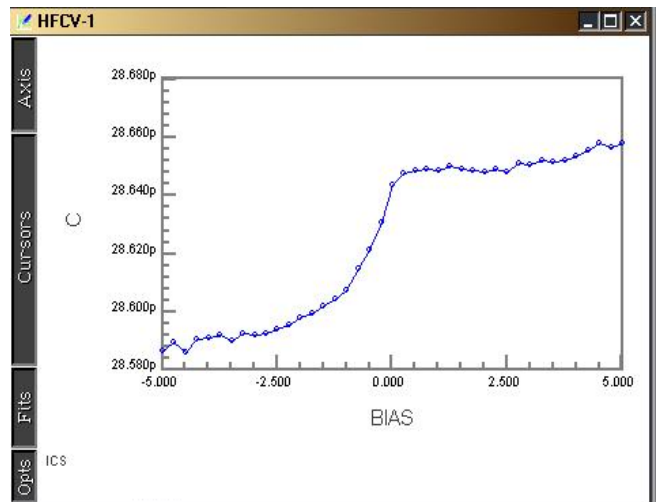

Fig. 4. The relations between oxide capacitor in MOS diode and voltage.

From figure 4 when voltage send to capacitor increase to 1 V, then the result of capacitance is 28.64 pF, and when increase voltage to capacitor is 4 V then the result of capacitance is will a little alteration to 28.67 pF, so it able to capacitance estimation between aluminum metal layer and $p^+$ semiconductor of MOS diode at 28.62 pF

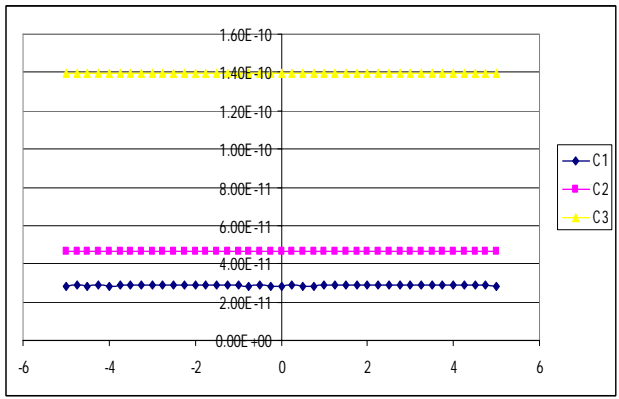

Fig. 5. The relations between oxide capacitor at each level oxide thickness in MOS diode and voltage.

From figure 5 in the testing we change silicon dioxide thickness in MOS diode that three period, 1st period ($C_1$) oxide thickness layer at 150 nanometer, than capacitance about 140 pF, 2nd period ($C_2$) oxide thickness layer at 300 nanometer, then capacitance about 47 pF, 3rd ($C_3$) oxide thickness layer at 500 nanometer, then capacitance about 28.2 pF, this the best period as figure 4.

## 5 THE CAPACITANCE AND RELATIONSHIP OF SILICON DIOXIDE BETWEEN ALUMINUM METAL AND N+ SEMICONDUCTOR

When send voltage increase from -5 V until to 5 V and measurement the capacitance and voltage (C-V) between aluminum metal layer and $n^+$ semiconductor at capacitance alteration follow on voltage. From graph when voltage is send to C increasing follow on the capacitance result, and the capacitance result from measurement, it has be similar to each result. For example from figure 6 when voltage is sent to capacitor increasing at 1 V then the capacitance is 29.52 pF

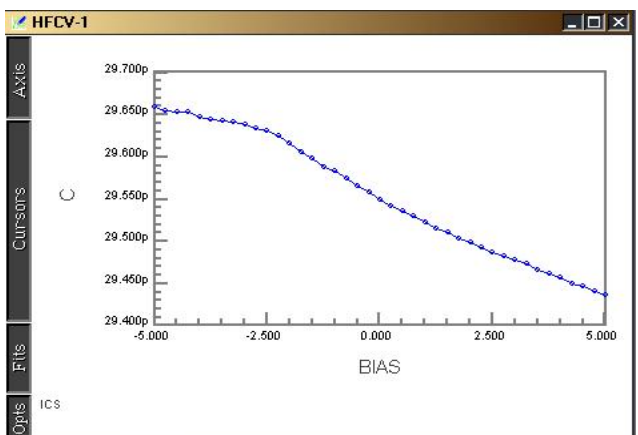

Fig. 6. The relations between oxide capacitor in MOS diode and voltage.

When increased voltage to capacitor at 4 V, than the capacitance is little alteration 29.47 pF, thus it able to estimate capacitance between aluminum metal layer and $n^+$ semiconductor of MOS diode it has around 29.55 pF.

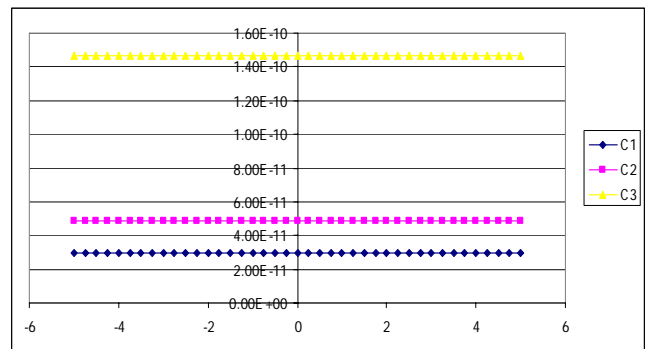

Fig. 7. The relations between oxide capacitor at each layer oxide thickness in MOS diode and voltage.

From figure 7 in the testing we change silicon dioxide thickness in MOS diode that three period, 1st period ($C_1$) oxide thickness layer at 150 nanometer, than capacitance about 140 pF, 2nd period ($C_2$) oxide thickness layer at 300 nanometer, then capacitance about 47 pF, 3rd period ($C_3$) oxide thickness layer at 500 nanometer, then capacitance about 28.2 pF, this the best period as figure 6.

## 6 THE CAPACITANCE AND RELATIONSHIP OF SILICON DIOXIDE BETWEEN ALUMINUM METAL LAYER 1 AND ALUMINUMS METAL LAYER 2

In this testing has designed and establish the structure of metal layer 1 and metal layer 2 for finding the capacitance of oxide capacitor between both of two layers. Evaluation and testing by changing the silicon dioxide thickness in MOS diode that 3 periods, the 1st period ($C_1$) oxide layer thickness at 150 nanometer, then the capacitance around 82 pF, 2nd period ($C_2$) oxide layer thickness at 300 nanometer, then the capacitance around 27 pF, 3rd period ($C_3$) oxide thickness at 300 nanometer, then the capacitance around 16 pF, this the best capacitance.

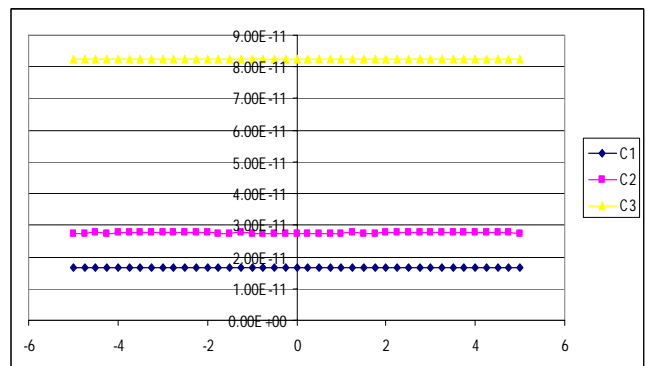

Fig. 8. The relations between oxide capacitor at each of oxide thickness in 1st metal layer and 2nd metal layer.

From the result, the best oxide capacitor of both metal layer around 16 pF at silicon dioxide thickness 500 nanometer and it the best thickness, because of it made capacitance is a little, and silicon dioxide thickness has be function to oxide insulator and protect impure substance too.

## 7 CONCLUSION

From the result, when we designing and establish MOS diode structure for find out the relations of oxide capacitor at suitable for establish CMOS in the integrated circuit level 5 micrometer technology. Then it able to know about the best level silicon dioxide thickness is 500 nanometer, after that the suitable capacitance in the integrated circuit and decrease problem from CMOS effect when use in low frequency.

## APPENDIX

Capacitance and silicon dioxide relations between aluminum metal with $p^+$ semiconductor

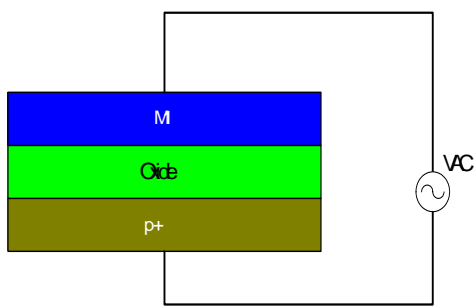

Capacitance and relationship of silicon dioxide between aluminum metal and $n^+$ semiconductor

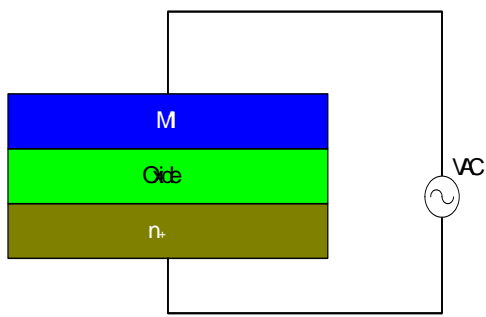

## ACKNOWLEDGMENT

The researchers, we are thank you very much to our parents, who has supporting everything to us. Thankfully to all of professor for knowledge and a consultant, thank you to Miss Suphansa Kansa-Ard for her time and supporting to this research. The last one we couldn't forget that is Kasem Bundit University, Engineering Faculty for supporting and give opportunity to our to development in knowledge and research, so we are special thanks for everything.